\documentclass[twocolumn,prl,showpacs,floatfix]{revtex4}
\usepackage[final]{graphicx}
\usepackage[usenames]{color}
\usepackage{afterpage}
\usepackage{bm}

\date{\today}
\begin{document}

\title{Strong charge-transfer excitonic effects and Bose-Einstein exciton-condensate in graphane}

\author{Pierluigi Cudazzo$^1$, Claudio Attacalite$^1$, Ilya V. Tokatly$^{1,2}$
and Angel Rubio$^{1,3}$}

\affiliation{$^1$ Nano-Bio Spectroscopy group and ETSF Scientific Development Centre, 
  Dpto. F\'isica de Materiales, Universidad del Pa\'is Vasco, 
  Centro de F\'isica de Materiales CSIC-UPV/EHU-MPC and DIPC, 
  Av. Tolosa 72, E-20018 San Sebasti\'an, Spain\\ 
  $^2$ IKERBASQUE, Basque Foundation for Science, E-48011 Bilbao, Spain\\
  $^3$ Fritz-Haber-Institut der Max-Planck-Gesellschaft, 
  Theory Department, Faradayweg 4-6, D-14195 Berlin-Dahlem, Germany}

\begin{abstract}
Using first principles many-body theory methods (GW+BSE) we demonstrate that
optical properties of graphane are dominated by localized charge-transfer
excitations governed by enhanced electron correlations in a two-dimensional
dielectric medium. Strong electron-hole interaction leads to the appearance of
small radius bound excitons with spatially separated electron and hole, which
are localized out-of-plane and in-plane, respectively. The presence of such
bound excitons opens the path on excitonic Bose-Einstein condensate in graphane
that can be observed experimentally.
\end{abstract}


\maketitle

\begin{figure}
\includegraphics[clip,width=1.0\linewidth]{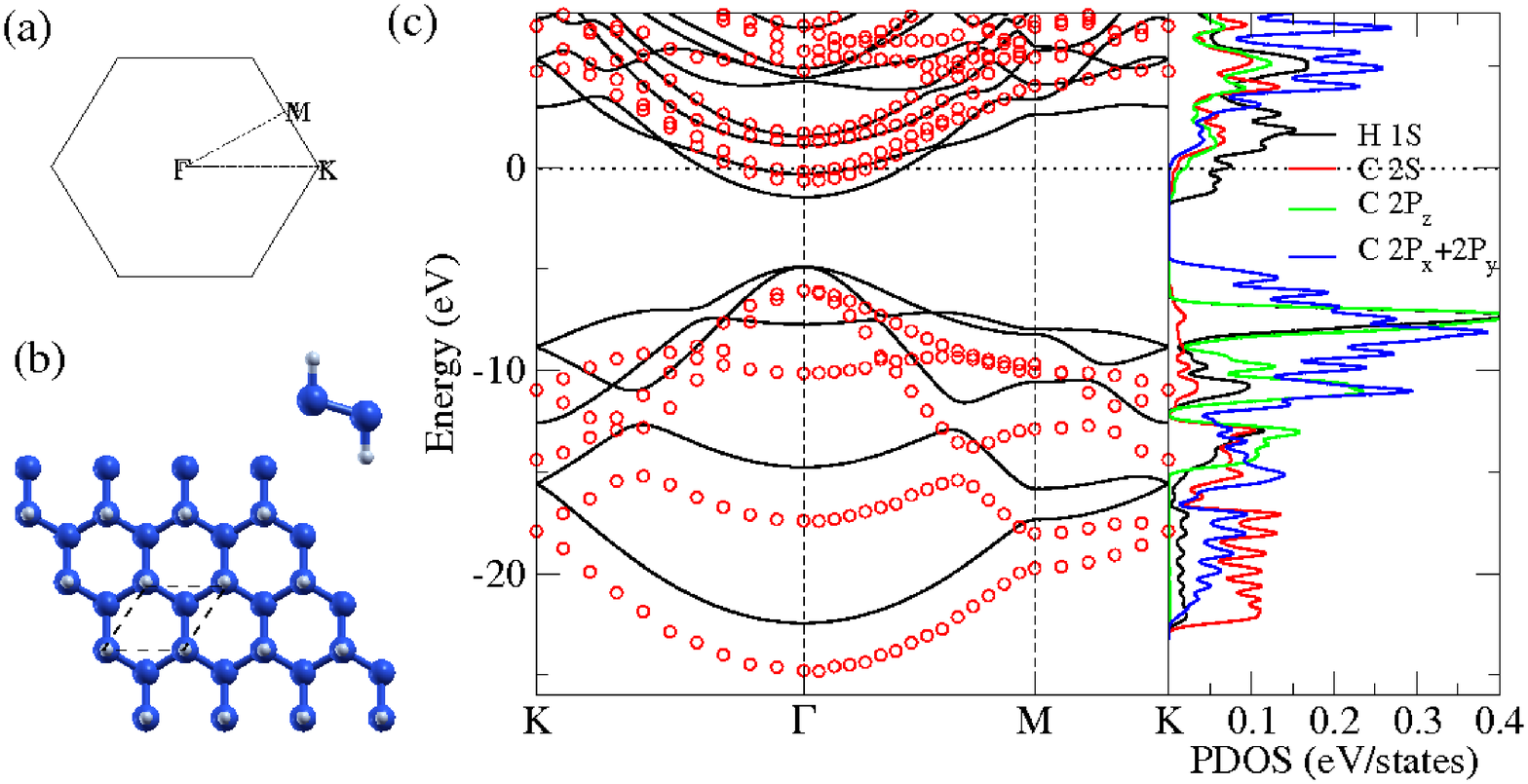}
\caption{First Brillouin zone (a) and
unit cell and bases of graphane in the chair conformation (b). Blue and white
balls represent carbon and hydrogen atoms respectively.
(c) Band structure in LDA (full line) and GW approximation
(red circles) and projected density of states of graphane. The zero
indicates the vacuum level position.} 
\label{bande}
\end{figure}

Despite the short-live of graphene\cite{novoselov}
and its derivatives the understanding and control of their
properties rapidly approach a maturity\cite{geim1,geim}. Chemical modifications by
oxidation\cite{ruoff}, functionalization and doping\cite{dai} have enhanced
foreseen applications in nanotechnology.
Recent synthesis of the fully hydrogenated-graphene (named "graphane")\cite{elias}, which has been predicted to be a wide band-gap insulator with gap of 5.4 eV\cite{sofo,lebegue}, adds to the portfolio of
carbon-based structures for nanodevice applications~\cite{sofo,geim}. The stability
of the new 2D material has been analysed and two geometries (chair and boat) have
been proposed\cite{sofo}. However, there are still few open questions that need
to be addressed: $i$) how much H is really incorporated in the samples? $ii$) 
does graphane inherit negative electron affinity of hydrogenated diamond samples
used for electron-emitters? $iii$) what is the role of electron correlations in
the band-structure and screening; are electron-hole effects as important as for
carbon nanotubes\cite{louie1,louie2,chang} or small as in diamond\cite{marini}?
The present letter address those issues, in particular the last three, by means
of first-principles calculations based on many-body Green function theory.
Our results indicate the possibility of having an excitonic Bose-Einstein
condensate (BEC) 
by continuous pumping of excitons by light. Similar to the recent induced
photoluminescence in graphene oxygenation\cite{gokus}, the strong spatial
localization of the excitons in graphane points to a highly efficient
defect-induced luminescence in chemically (or doped) modified graphane.
In our ground state calculations of the most stable chair conformation of graphane
(Fig.\ref{bande} (b)) we used a pseudopotential plane-wave
approach\cite{pwscf,suppl} within the LDA approximation to DFT\cite{sham}. While
excited states have been treated using the state-of-the-art many body
approach\cite{suppl,hedin2,onida,yambo}. 

Before addressing the excitonic effects in the optical properties of graphane
including the formation of BEC, we emphasize that although the
dynamical stability of the ideal graphane in the chair conformation
(Fig.\ref{bande} (b)) has been recently shown\cite{ciraci}, the calculated
phonon spectra agrees only partially with the measured Raman data\cite{elias}. In
particular the measured $D$ peak agrees well with the calculated Raman active
E$_g$ phonon at 1340 cm$^{-1}$ while the $D^{\prime}$ at 1620 cm$^{-1}$ most
likely originates from impurities as H vacancies. This means that there is a
large defect density in graphane making it a hole-doped semiconductor with H
vacancies acting as center for induced luminescence (see below). However at low
doping H vacancies do not affect significantly the electronic structure, so that
ideal graphane allows to obtain a realistic description of the electronic and
optical properties of this system. On the other hand, the stability of graphane
as hole-doped semiconductor, suggests the possibility to achieve a metallic
phase under strong hole-doping by dehydrogenation. Under this condition we have
observed a strong electron-phonon renormalization of the phonon
frequencies around the BZ center (Kohn-anomaly)\cite{suppl}. Similar to the case
of doped diamond\cite{pickett}, this is a manifestation of a strong
electron-phonon coupling pointing to the possibility of a high T$_c$
superconducting phase.  
         
{\it Electronic properties:} the  band structure of graphane (Fig.\ref{bande} (c))
is mainly dictated by the sp$^3$ hybridization of carbon orbitals 
(see Fig. \ref{bande} (b)). This
causes a band gap opening (3.4~eV at the DFT-Kohn-Sham level) with respect to
graphene.
The classification of states is especially simple at the $\Gamma$-point. The top/bottom valence(E$_g$)/(A$_{1g}$) corresponds to the C-C bonding $\sigma$ states, while two intermediate 
occupied bands corresponds to C-H  bonding $\sigma$ states ($A_{1g}$). The
bottom of the conduction band is the anti-bonding A$_{2u}$ C-H $\sigma$ state and the anti-bonding E$_u$ C-C $\sigma$ state appears at higher energy
(4.38 eV). 
The other bands correspond to free-electron-like states. Electron self-energy
effects computed at the $GW$-level\cite{suppl,hedin2,yambo}  
modify strongly the band structure (red dots on Fig.\ref{bande} (c)): the fundamental energy gap at $\Gamma$ increases from 3.4 eV to 5.4 eV, while the gaps at the high symmetry points M and K become 
14.2 eV and 15.9 eV, instead of LDA values of 10.6 eV and 11.8 eV, in agreement
with recent calculations\cite{lebegue}. Moreover we found that the vacuum level
is at about 4.9 eV from the top of the LDA valence band (see Fig.\ref{bande}
(c)) so that graphane has a positive electron affinity both in LDA (1.44 eV) and
GW (0.66 eV).

\begin{figure}
\includegraphics[clip= ,width=1.0\linewidth]{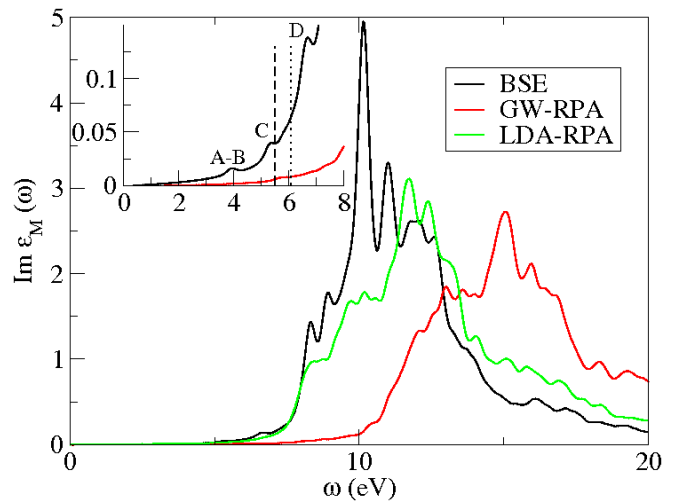}
\caption{Imaginary part of the macroscopic dielectric function for light
polarized along the graphane plane. The dashed and dotted lines in the inset indicate the
position of the GW gap and vacuum level respectively. For light propagating
along the $z$ axes the spectrum (not shown) is flat up to 6 eV.}
\label{epsilon}
\end{figure}

{\it Optical properties of graphane}: In Fig.\ref{epsilon} we plot the imaginary part of the macroscopic dielectric
function $\epsilon_M(\omega)$ for a light propagating along the graphane plane
in the $x$-direction. 
To reveal the physical origin of different features in the optical spectrum
we compare $\rm{Im}\epsilon_M(\omega)$ calculated
 ($i$) without taking into account both the inter-electron and the
electron-hole correlations, LDA-RPA, that is the random phase approximation (RPA) on top of the bare LDA band
structure, ($ii$)  RPA using the  $GW$ quasi-particle spectra (GW-RPA) 
which neglects electron-hole correlations, and, finally, ($iii$) from the full
solution of the Bethe Salpeter equation (BSE) which accounts the
excitonic effects\cite{suppl,yambo,onida}. 

The LDA-RPA absorption spectrum (green line in Fig.\ref{epsilon}) do not show any
significant feature near to the LDA band gap (3.4 eV) because the corresponding transitions, being dipole-allowed, have small oscillator strength as the overlap between the top-valence (localized on the C-C bond,) 
and bottom-conduction (localized on the C-H bond) states is small. Pronounced features in $\epsilon_M(\omega)$ are present at higher
energies. The peak about 8.5 eV is related to vertical
transitions from the valence band to the conduction band at large wave
vectors, while the peak at 10.2 eV corresponds to transitions near to the M point. The large structure at 11.2 eV arise
from a Van Hove singularity  near $M$ corresponding to transitions from the
valence band to out-of-plane states with Kohn-Sham energies between 3.37 eV and
4.14 eV. 
The main effect of the quasi-particle corrections is a strong global shift of the absorption spectrum to higher energies (GW-RPA curve in Fig.\ref{epsilon}).
Since the optical transition energy increases the amplitude of $\epsilon_M(\omega)$ is reduced to satisfy the f-sum rule. 
However, we show that electron-hole correlations dramatically modify the shape
of $\epsilon_M(\omega)$ (see BSE curve in Fig.\ref{epsilon}). 
The ``bulk'' of the absorption spectrum nearly comes back to the original LDA-RPA position, 
but with a significant redistribution of the oscillator strengths from the higher to the lower energies, 
and an appearance of a number of pronounced excitonic resonances that are
related to electron-hole pairs with electrons in the delocalized out-of-plane
states\cite{suppl}.

\begin{figure}
\includegraphics[clip= ,width=0.9\linewidth]{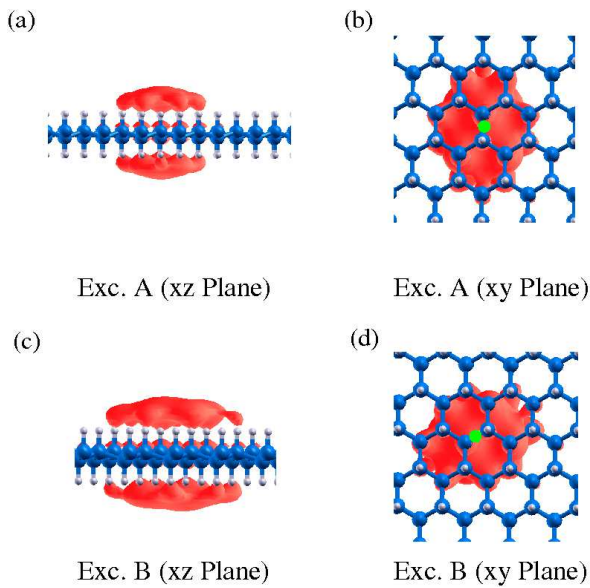}
\caption{3D-Shape of the low energy excitons wave function for fixed position of
the hole (green circle).} 
\label{exciton}
\end{figure}

In this work we concentrate on the most prominent physical effect of the electron-hole interactions 
in graphane, the appearance of bound excitons below the GW gap (see inset in Fig.\ref{epsilon}). 
These excitations, which are completely missing in RPA, are responsible for the
UV absorption of graphane. The absorption spectrum for light propagating along
$x$ shows two bound excitons ($B$ and $C$) with large binding energies of 1.6~eV
and 0.3~eV, respectively. The solution of the BS equation also reveals the
existence of a dark exciton ($A$), 2~meV below the first optically active $B$-exciton. 
The exciton $C$ is related to transitions from the highest valence band to the
first out-of-plane band and its wave function is delocalized
out-of-plane\cite{suppl}.
In contrast, the spatial extension of the strongly bound excitons $A$ and $B$ is very small,
$r_{ex}\approx 5.0$~\AA, as it can be seen in Fig.\ref{exciton} (b) and (d). These excitons are formed from
the states of the double degenerate (at $\Gamma$-point) E$_{g}$ valence band and the A$_{2u}$ conduction band (the hole states in these excitons belong to the lowest, E$^{(2)}_{g}$, and the highest, E$^{(1)}_{g}$, bands respectively).
Their wave functions reflect the C-H anti-bonding character of the states with
the electron mainly localized on top of the H-atoms.
Since the hole is localized on the C-C bonds, the creation of such
excitons corresponds to a charge transfer from the middle of the carbon plane to the
side planes on top of the hydrogen atoms (see Fig.\ref{exciton} (a)
and (c)).

It is instructive to compare absorption spectra of graphene\cite{louie_g} and graphane. Electron-hole
interaction is important in both systems. However, in the former it
leads only to a redistribution of the oscillator strength resulting in
the appearance of a resonant exciton at 4.5 eV, while in the latter it gives rise to a
strongly bound exciton at 3.8 eV where the spectrum of graphene is completely
flat. This can be used as an additional experimental fingerprint
of the presence of hydrogenated platelets embedded in graphene.

Considering the light propagating in the $y$-direction we find that the role of the strongly bound excitons
is inverted: $A$ exciton becomes optical active, while $B$ exciton
is dark. Since the two excitons are nearly degenerate and have identical oscillator strengths for the
corresponding polarization of light the absorption spectra remain almost unchanged. The reason for this is
the symmetry of Bloch states involved in the formation of the excitons: the electronic states have A$_{2u}$
symmetry, while the hole states belong to the 2D irreducible representation E$_g$ with the eigenfunctions
$\psi^{(1)}_{E_g}$ and $\psi^{(2)}_{E_g}$. A group theoretical
analysis\cite{suppl} shows
that the only nontrivial dipole matrix elements are: $\langle\psi_{A_{2u}}|x|\psi^{(1)}_{E_g}\rangle$ and
$\langle\psi_{A_{2u}}|y|\psi^{(2)}_{E_g}\rangle$, which explains why the $A$ and $B$ excitons are only
visible for the $y$- and $x$-polarization, respectively. For the light propagating perpendicular to the
graphene plane both excitons are dipole forbidden.

The presence of H-vacancies with the strong spatial localization of the lowest
$A$-exciton makes graphane to luminescence with high yield upon UV excitations,
similarly to layered h-BN where a high luminescence yield is observed from a
strong localized exciton\cite{watanabe}.

A striking feature of the excitons $A$ and $B$ is their large binding energy, which at least by an order of magnitude exceeds that in typical wide gap semiconductors. 
This becomes especially surprising if we note that both the A$_{2u}$ conduction band and the E$_g$ valence bands are nearly perfect parabolas over a large fraction of BZ, 
and, therefore, the excitons, despite their small radius, should be reasonably
well described within the effective mass approximation. Using the standard
method of invariants \cite{BirPikus} we derived the $\mathbf{k}\cdot\mathbf{p}$
Hamiltonian at $\Gamma$ for the valence bands
\begin{equation} \label{eq1}
\hat{H}(\mathbf{k})=\alpha\hat{I}(k^2_x+k^2_y)+\beta[\hat{\sigma}_z(k^2_x-k^2_y)+2\hat{\sigma}_xk_xk_y],
\end{equation}
and for the conduction band
$\hat{H}(\mathbf{k})=\frac{k^2_x+k^2_y}{2m_e}$\cite{suppl}, from which
we find the value of the reduced electron-hole mass $\mu_{ex}=0.29m_0$, which is very typical for most known semiconductors 
(here $m_0$ is the bare electron mass, and for simplicity we model the two valence bands by a single parabola with an average mass). A resolution of the apparent paradox is an unusual form of the effective electron-hole interaction in graphane.
In 3D dielectric media the effective interaction is obtained by the replacement $e^2/r\to e^2/\epsilon r$, where $\epsilon$ is the static dielectric
constant. This simple recipe does not work in graphane that is a 2D dielectric, where the very notion of the dielectric constant makes no sense.
In fact, an external electric field $E$ induces a polarization $P$ which is bound to the plane: $P({\bf r}) = \alpha_{2D}E({\bf r})\delta(z)$, where
$\alpha_{2D}$ is the internal polarizability of the 2D dielectric. Using the above relation in the Poisson equation for a point charge we get the
following effective interaction potential (its 2D Fourier component)

\begin{equation} \label{Veff}
V_{eff}(\mathbf{q})=\frac{2\pi e^2}{|\mathbf{q}|(1+2\pi\alpha_{2D}|\mathbf{q}|)}
\end{equation}

which is very different from the trivial renormalization of charge in 3D
systems. The only parameter entering Eq.\ref{Veff} is the polarizability
$\alpha_{2D}$ of a single graphane layer, which we extract from our {\it ab
initio} results. Since our calculation are performed for a periodic stack of
layers with sufficiently large inter-layer distance $L$, the 2D polarizability is related to the actually calculated 3D polarizability as follows
$\alpha_{2D}=L\frac{\epsilon-1}{4\pi}$, where $\epsilon$ is the dielectric
constant of our 3D multilayer system.
We can also estimate the exciton binding energy $E_{ex}$ by the variational solution
of the Schr\"odinger equation with the effective potential in Eq.\ref{Veff} using the simplest trial wave function,
$\psi\sim e^{-r/a}$, where $a$ is the variational parameter\cite{prb}. The result $E_b=2.0$~eV is very close to the ab-initio value of 1.6~eV, while the optimized value of the
variational parameter $a=7.7$a.u. almost perfectly matches the actual excitonic radius.
Hence we clearly see that unusual strong binding of excitons
in graphane is a result of a weak and specifically nonlocal 2D dielectric screening. 

Because of the small radius and the large binding energy of excitons it is tempting to consider graphane as a potential candidate for a realization of the Bose-Einstein condensation (BEC) of optically pumped excitons. Indeed, using the standard expression for the degeneracy temperature $T_0$ of a 2D Bose gas\cite{bse2d}:

\begin{equation}
k_BT_0=\frac{2\pi\hbar^2 n}{M_{ex}}
\end{equation}

(here $M_{ex}=m_c + m_v\approx 1.3m_0$ is the excitonic mass) we find that at room temperature $T_0=300K$ the quantum degeneracy of excitonic gas in
graphane is reached for the density $n\approx 5$~$10^{12}$cm$^{-2}$, which is two orders of magnitude smaller than $\frac{1}{\pi a_{ex}^2}=5.1$~$10^{14}$cm$^{-2}$
(the later value sets an order of magnitude of the Mott critical density for the transition from the gas of excitons to a gas/liquid of unbound electrons
and holes). Another argument in favor of excitonic BEC in graphane is a spatial separation of the charges in the excitonic state -- the hole is localized
in the middle C-layer, while the electron on top of the H-layers. Therefore graphane looks quite similar to the coupled quantum wells (CQW) structures,
which are very popular for experiments on excitonic BEC \cite{nature1,nature2}.  
In CQW \cite{nature1,nature2} such a separation leads to a very long excitonic life time that reaches microseconds. To estimate the radiative life time
$\tau$ of the excitons in graphane we adopted a 2D version of the approach of Ref.\cite{lui}, which lead to the following relation

\begin{equation}\label{lifetime}
\frac{1}{\tau}=\frac{4\pi e^2\Omega}{\hbar c}\frac{d^2}{A},
\end{equation}

where $\Omega$, $d$ and $A$ are the exciton frequency, dipole matrix element, and the unit cell area, respectively. The result we find, $\tau=15$ ps, is
not very encouraging -- the value is too small to unconditionally speak about BEC.
On the other hand, it is still larger that the lifetime of cavity
polaritons where BEC has probably been observed \cite{nature3}. Clearly a detailed analysis of the thermalization kinetics is needed to make more
definite predictions. Defects will also play a role in pinning the exciton
condensate. Nonetheless, we believe that BEC of strongly bound, small radius excitons in graphane is an interesting possibility which deserves
to be carefully studied both experimentally and theoretically.

This work was supported by
the Spanish MEC (FIS2007-65702-C02-01),
ACI-Promociona (ACI2009-1036) Grupos Consolidados UPV/EHU del Gobierno
Vasco" (IT-319-07), the European Union through e-I3 ETSF project (Contract Number
211956). We acknowledge support by the Barcelona Supercomputing Center, Red Espanola
de Supercomputacion".

\clearpage



 




\begin{center}
\large{\bf Supplementary Material}
\end{center}

\section{Numerical details}

Our calculations of the ground state properties are based on density functional
theory\cite{hohemb,sham-s} (DFT) within local density approximation\cite{perdew} (LDA)
implemented in the pseudo-potential plane wave framework\cite{pwscf-s}. The chair
conformation has been simulated using a two
dimensional triangular lattice with a basis of two CH units where the ionic
potential has been modeled using Bachelet, Hamman and Schl\"uter
pseudopotentials\cite{bach}. The distance between two adjacent layers has been set to
15\AA~ to simulate an isolated Graphane sheet. We chose a 90 Ry kinetic energy
cutoff for the electronic wave function expansion and a $16\times 16\times 1$
Monkhorst-Pack\cite{pack} mesh for Brillouin zone (BZ) integration. Then, the optimized unit
cell has been obtained minimizing the total energy as a function of the lattice
parameter. At each value of the lattice constant the atomic
positions ($i.e.$ the internal degree of freedom) were fully relaxed to minimize
the Hellman-Feynman\cite{feynman} forces. In the optimized structure the two CH 
units relax up-word and down-word respect to the ideal position of the graphene
layer resulting in the formation of sp$^3$ C-C and C-H bonds. In particular the
bond length are 1.51 \AA~ (for the C-C bonds) and 1.11 \AA~ (for the C-H bonds)
in agreement with previous calculations\cite{sofo-s,bouk}. 

Phonon frequencies were calculated using the plane waves method within Linear
Response Theory\cite{giannozzi} using a $30\times 30\times 1$ Monkhorst-Pack $k$-point mesh to
compute the dynamical matrix. This allows us to achieve a very good convergence
on the phonon spectrum (about 2\% in the phonon frequency). A $10\times 10\times
1$ $q$-mesh in the BZ was used to interpolate the force constants for the phonon
dispersion calculation.

Finally in calculating the QP energies we used the GW approximation of
Hedin\cite{hedin2-s,yambo-s}, and the Bethe-Salpeter equation (BSE) has been solved
to cope with excitonic effects in the absorption spectra\cite{onida-s,yambo-s}. In
particular in our calculation of QP energies we used a $30\times 30\times 1$
$k$-point mesh and 100 bands for computing the self energy. We take into account
dynamical screening effects in W through the generalized plasmon pole model. 
The same $k$-point mesh with 40 bands has been used to evaluate the dielectric
function in RPA. 
The electron-hole kernel in the BSE and the dielectric function has been
evaluated on a $30\times 30\times 1$ k-point mesh. Moreover we used 100 bands
for the calculation of $W$ and 20 bands to evaluate $\epsilon_M(\omega)$ since we
were interested to the low energy region of the spectrum.

\section{Phonon spectra for pure and doped graphane}

\begin{figure}
\includegraphics[clip= ,width=1.0\linewidth]{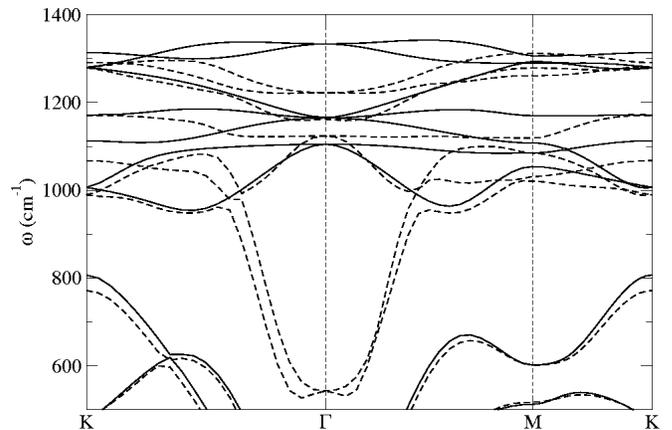}
\caption{Phonon spectra of ideal (solid line) and 1$\%$ hole-doped (dasched line)
graphane in the chair conformation}. 
\label{ph}
\end{figure}

In Fig.\ref{ph} (solid line) we show the calculated phonon dispersion
curve for a perfect graphane in the chair configuration. The phonon spectra
presents three well separated types of vibrations: the low-frequency acoustic
modes (not shown), the intermediate-frequency optical modes involving the
stretching of the C-C bonds, and separated by a gap of 1500 cm$^{-1}$ (not
shown), we find optical modes corresponding to
C-H bonds. The presence of this large gap proves that the vibrations of H atoms
are almost completely decoupled from the other modes.

Comparison of our calculated spectra with available experimental Raman data,
shows that the measured $D$ peak agrees well with the calculated Raman active
E$_g$ phonon at 1340 cm$^{-1}$ while the $D^{\prime}$ peak observed at 1500
cm$^{-1}$ most likely is related to impurity effects due to H vacancies.

Finally, in Fig.\ref{ph} we compare the phonon spectra of ideal graphane (solid
line) with that of $1\%$ hole-doped graphane (dashed line) obtained in the
rigid-band approximation. As we can see hole-doping causes a strong
renormalization of the phonon spectra near to the $\Gamma$ point. In particular
the two degenerate optical modes related to the C-C stretching with H atoms
moving in-phase with C atoms soften from 1168 cm$^{-1}$ to 542 cm$^{-1}$ while
the two degenerate optical modes related to C-C stretching with H atoms moving
out-of-phase with C atoms soften from 1333 cm$^{-1}$ to 1221 cm$^{-1}$. The
large Kohn anomaly at 542 cm$^{-1}$ suggest a strong electron-phonon coupling in
doped graphane related to the C-C covalent bonds. This makes doped graphane a
possible candidate for high T$_c$ superconductivity.

\section{Symmetry properties of the wave functions}

In this section we perform an analysis of symmetry properties of the wave
functions corresponding to the valence and conduction bands at the $\Gamma$ point
based on the group theory. This allows to explain the behaviour of the
absorption spectra at low energy for different polarizations and the nature of
the two nearly degenerate excitons.  

Graphane in the chair conformation belongs to the $D_{3d}$ point group which is
characterized by 12 symmetry operations (with inversion) and 6 irreducible
representations: two 2D representations E$_g$ and E$_u$ and four 1D
representations A$_{1g}$, A$_{2g}$, A$_{1u}$ and A$_{2u}$ (see table \ref{tabd3d}).

\begin{table}[hc]
\begin{center}
\begin{tabular}{|c|c|c|c|c|c|c|c|}
\hline
& $E$ &  $2C_3$  &  $3C^{\prime}_2$  &
$i$ & $2S_6$ & $3\sigma_d$ & Basis functions  \\  
\hline
$A_{1g}$ & 1 & 1 & 1 & 1 & 1 & 1 & $x^2+y^2$, $z^2$ \\
\hline
$A_{2g}$ & 1 & 1 & -1 & 1 & 1 & -1 & \\
\hline
$E_g$ & 2 & -1 & 0 & 2 & -1 & 0 & $(x^2-y^2,xy)$ $(xz,yz)$ \\
\hline
$A_{1u}$ & 1 & 1 & 1 & -1 & -1 & -1 & \\
\hline
$A_{2u}$ & 1 & 1 & -1 & -1 & -1 & 1 & z \\
\hline
$E_u$ & 2 & -1 & 0 & -2 & 1 & 0 & $(x,y)$ \\  
\hline
\end{tabular}
\end{center}
\caption{Character table for D$_{3d}$ point group.}
\label{tabd3d}
\end{table}

\begin{figure}
\includegraphics[clip= ,width=1.0\linewidth]{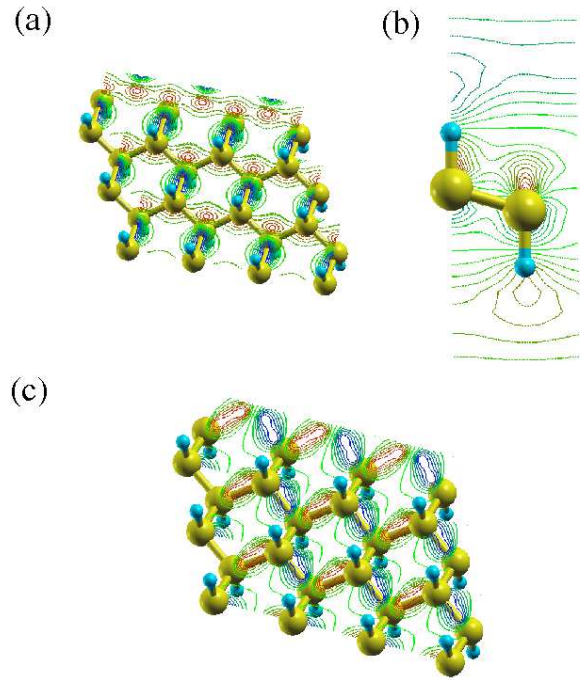}
\caption{Projection on the $XY$ ((a) and (c)) and $ZY$ (b) planes of the wave
functions $\psi^{(2)}_{E_{2g}}(\mathbf{r})$ (a), $\psi_{A_{2u}}(\mathbf{r})$ (b)
and  $\psi^{(1)}_{E_{2g}}(\mathbf{r})$ (c). The red-blue lines refer
to negative-positive values of the wave function.}
\label{charge}
\end{figure}

Since the valence bands are two fold degenerate at the $\Gamma$ point they
belong to one of the 2D irreducible representation E$_g$ or E$_u$. However as
can be inferred from Fig.\ref{charge} (a) and (c), the wave functions corresponding to the valence
bands are symmetric respect to the inversion. This is sufficient to conclude
that the valence bands indicated with $\epsilon^{(1)}_{E_g}$ and with
$\epsilon^{(2)}_{E_g}$ belong to
the E$_g$ irreducible representation. Moreover we indicate with
$\psi^{(1)}_{E_g}$ and $\psi^{(2)}_{E_g}$ the corresponding wave function, with
the label 1(2) indicating that the wave function is antisymmetric (symmetric)
under reflection respect to the $YZ$ plane.

The conduction band is not degenerate and it is characterized by a wave function
antisymmetric respect to the inversion (see Fig.\ref{charge} (b)). On the other hand this latter is
antisymmetric under refection respect to the $YZ$ plane. Thus the conduction
band ($\epsilon_{A_{2u}}$) belongs to the A$_{2u}$ (see table\ref{tabd3d}) irreducible representation.

Now we focus on the symmetry properties of a general polar vector ($x,y,z$) as for example
the dipole moment. It can be easily shown that the third component $z$
transforms in agreement with the A$_{2u}$ irreducible representation, while the
in-plane components ($x,y$) transform in agreement with the E$_u$ irreducible
representation. In particular $y$ is the partner of $x$. 
Then, when the light is directed along the $Z$ axis the possible states coupled
with the valence bands by dipole moment are contained in the direct product
$A_{2u}\times E_g=E_u$. This means that transitions from the valence to the
conduction bands are forbidden for light directed along the $Z$ axis and both
excitons $A$ and $B$ are not visible. On the other hand, when the light is
directed along $X$ or $Y$, the possible states coupled with the valence bands by
dipole moment are contained in the direct product     
$E_u\times
E_g=A_{1u}+A_{2u}+E_u$ which is a 4D reducible representation of the symmetry
group in the bases defined by the four wave functions: $x\psi^{(1)}_{E_g}$,
$y\psi^{(1)}_{E_g}$, $x\psi^{(2)}_{E_g}$ and $y\psi^{(2)}_{E_g}$. It is now
possible, starting from this bases set, to generate a set of wave functions
symmetrically adapted with the three irreducible representation appearing in the
direct product $E_u\times E_g$. In particular we found that the only wave
function which transform in agreement with the A$_{2u}$ irreducible
representation is a linear combination of $x\psi^{(2)}_{E_g}$ and
$y\psi^{(1)}_{E_g}$.
This means that the only possible transitions involving the wave function
$\psi_{A_{2u}}$ are: $\langle\psi_{A_{2u}}|x|\psi^{(1)}_{E_g}\rangle$ and
$\langle\psi_{A_{2u}}|y|\psi^{(2)}_{E_g}\rangle$. 
     
It is now
clear why the most bounded exciton related to electron-hole pairs
($\psi_{A_{2u}},\psi^{(2)}_{E_{2g}}$) is optical active for light polarized
along $Y$, while the highest energy exciton related to electron-hole pairs
($\psi_{A_{2u}},\psi^{(1)}_{E_{2g}}$) is visible for light polarized along $X$. However, due to the
small overlap between $\psi_{A_{2u}}$ and $\psi_{E_{g}}$ wave functions
the two excitons present a very small strength compared with that of the
higher energy excitons. As a results, it does not reflect with evident features
on the behaviour of $\epsilon_M(\omega)$.

\section{Analysis of the excitonic resonances in the graphane absorption spectra
of Fig.2}

\begin{figure}[hc]
\includegraphics[clip= ,width=1.0\linewidth]{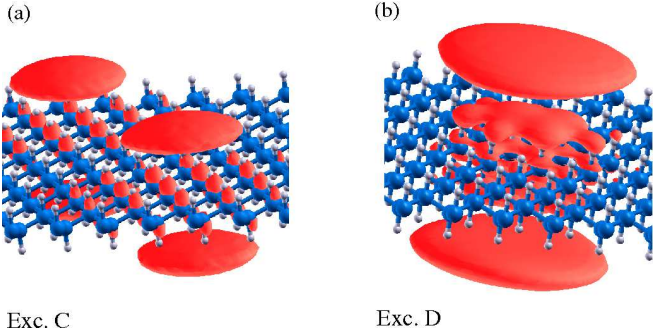}
\caption{3D-Shape of the excitonic wave function for the exciton C (a) and D
(b).}
\label{exciton-s}
\end{figure}

The prominent peak at 15.4 eV is completely depressed, while the absorption is
strongly enhanced at 10.2~eV and 11.0~eV. 
The first peak at 10.2~eV is related to the correlated electron and hole states with large wave vectors from the E$_g$ bands and out-of-plane bands 
with QP energies 0.9 eV $\le\epsilon^{QP}_{n\mathbf{k}}\le$ 2.1 eV. The 
``binding energy'' is about 0.5~eV. The second resonant exciton at 11.0 eV has a
lower binding energy (0.2 eV) and is formed by 
electron-hole pairs from C-H valence states and out-of-plane states with QP
energy close to 1.20 eV. 
The series of excitonic resonances between 8.30 eV and 8.90 eV 
are related to the electron-hole pairs corresponding to transitions from the
small wave vector E$_{g}$ states to the out-of-plane states with QP energies 
0.13 eV$\le\epsilon^{QP}_{n\mathbf{k}}\le$ 2.02 eV. 
Finally, the lowest  resonant excitonic peak at 6.46~eV (peak $D$ in the inset in
Fig.2) originates from 
the mixture of the transitions from the highest E$_{g}$ valence band to the
A$_{2u}$ conduction band and the first out-of-plane band. 
Because of the free-electron character of the electronic states involved in all these electron-hole resonances 
the corresponding excitonic wave functions are completely delocalized in the
out-of-plane regions (see Fig.\ref{exciton-s} (b)).

\section{$\mathbf{k}\cdot\mathbf{p}$ Hamiltonian}

In this section we derive the $\mathbf{k}\cdot\mathbf{p}$ Hamiltonian
$\hat{H}(\mathcal{K})$ for the valence bands around the $\Gamma$ point
($\mathcal{K}$ being a second order tensor operator which depends from the
components of the wave vector $\mathbf{k}$). To determine the number of the
independent parameters in $\hat{H}(\mathcal{K})$ we use the method of the
invariants\cite{winkler}, which consists to impose the condition:

\begin{equation}\label{eqinv}
\hat{D}^{E_g}(g)\hat{H}(g^{-1}\mathbf{\mathcal{K}})\hat{D}^{E_g^{-1}}(g)=\hat{H}(\mathbf{\mathcal{K}})  
\end{equation}

for each transformation $g$ of the symmetry group. $\hat{D}^{E_g}(g)$ being the 2D
matrix representing $g$ in the E$_g$ irreducible representation.

Since the matrix Eq.\ref{eqinv} represents $N^2=4$ equations for the elements
$\hat{H}_{ij}(\mathcal{K})$ of $\hat{H}(\mathcal{K})$, these constraints allow
one to construct $\hat{H}(\mathcal{K})$. 

To determine the correct
form of the Hamiltonian, we first note that the tensor operator $\mathcal{K}$
can be decomposed into two irreducible tensor operators, one
($\mathcal{K}^{A_{1g}}=k^2_x+k^2_y$) belonging to the A$_{1g}$ irreducible
representation and one ($\mathcal{K}^{E_g}=(k^2_x-k^2_y,2k_xk_y)$) belonging
to the E$_g$ irreducible representation ($2k_xk_y$ being the partner
of $k^2_x-k^2_y$). Furthermore, $\hat{H}(\mathcal{K})$ can be decomposed on a
complete set of linearly independent $2\times 2$ dimensional matrices  that
transform according to those irreducible representations which are contained in
the direct product $E_g\times E_g=E_g+A_{1g}+A_{2g}$. Starting from the four
linearly independent matrices $\hat{I}$ (identity matrix), $\hat{\sigma}_x$,
$\hat{\sigma}_y$ and $\hat{\sigma}_z$ (Pauli matrices) we can construct a set of
linearly independent matrices $\hat{\chi}^{\Gamma}_{1}$ symmetrically adapted with
the irreducible representation $\Gamma=E_g, A_{1g}, A_{2g}$. This can be done
in terms of the standard group theory using the relation:

\begin{equation}\label{eqg}
\hat{\chi}^{\Gamma}_1=\sum_g\hat{D}^{E_g}(g)\hat{\chi}\hat{D}^{E_g^{-1}}(g)D^{\Gamma}_{11}(g)
\end{equation}   

where $\hat{\chi}$ runs on the four matrices ($i.e.$ the identity and
the three Pauli matrices) and $D_{ij}^{\Gamma}(g)$ is the matrix which represents
$g$ in the irreducible representation $\Gamma$. Eq.\ref{eqg} allows to obtain
from $\hat{\chi}$ a matrix symmetrically adapted with the first row of
the representation $\Gamma$. The corresponding partners
$\hat{\chi}^{\Gamma}_i$ can be constructed from:

\begin{equation}
\hat{\chi}^{\Gamma}_i=\sum_g\hat{D}^{E_g}(g)\hat{\chi}^{\Gamma}_1\hat{D}^{E_g^{-1}}(g)D^{\Gamma}_{i1}(g)
\end{equation}

with $i\neq 1$. In this case we found that $\hat{I}$ and $\hat{\sigma}_y$
transform in agreement with $A_{1g}$ and $A_{2g}$ respectively. While
$\hat{\sigma}_z$ transforms in agreement with the first row of $E_g$ and
$\hat{\sigma}_x$ is its partner. To determine the matrices symmetrically
adapted with $E_g$ we chose $(k^2_x-k^2_y,2k_xk_y)$ as basis for the
construction of $\hat{D}^{E_g}(g)$.

At this point, for each irreducible representation $\Gamma$, we can construct the
corresponding invariant ($\mathcal{T}_{\Gamma}$) as a product of symmetrically
adapted matrices $\hat{\chi}^{\Gamma}_i$ and irreducible tensor
components $\mathcal{K}^{\Gamma}_i$:
$\hat{\mathcal{T}}^{\Gamma}=\sum_i\hat{\chi}^{\Gamma}_i\mathcal{K}^{\Gamma}_i$.

In particular we obtain two invariants:
$\hat{\mathcal{T}}^{A_{1g}}=\hat{I}(k^2_x+k^2_y)$ and
$\hat{\mathcal{T}}^{E_g}=\hat{\sigma}_z(k^2_x-k^2_y)+2\hat{\sigma}_xk_xk_y$ but
we do not find any invariant transforming with the $A_{2g}$ representation. This
happens because at the second order the tensor operator $\mathcal{K}$ does not
contain any term belonging to the $A_{2g}$ representation (see table
\ref{tabd3d}). On the other hand, in the presence of a magnetic field along the
$Z$ axes ($H_z$), one other term $gL_zH_z$ appears in the Hamiltonian ($L_z$
being the $Z$ component of the angular momentum) which results in a third tensor
operator related to the commutator $[k_x,k_y]$ and belonging to the $A_{2g}$
representation. As a consequence a third invariant
$\hat{\mathcal{T}}^{A_{2g}}\propto\hat{\sigma}_y[k_x,k_y]$ appears in the Hamiltonian. 

Thus, each liner combination of the two invariants $\hat{\mathcal{T}}^{A_{1g}}$ and
$\hat{\mathcal{T}}^{E_g}$ satisfies by construction Eq. \ref{eqinv}, so that the
correct form of the $\mathbf{k}\cdot\mathbf{p}$ Hamiltonian is:

\begin{equation} \label{eq1-s}
\hat{H}(\mathbf{k})=\alpha\hat{I}(k^2_x+k^2_y)+\beta[\hat{\sigma}_z(k^2_x-k^2_y)+2\hat{\sigma}_xk_xk_y]
\end{equation}
   
with two independent parameters $\alpha$ and $\beta$. Diagonalizing
Eq.\ref{eq1-s}
we find that the band dispersion in $\mathbf{k}$ space is spherically symmetric
with $E(\mathbf{k})=(\alpha\pm\beta)(k^2_x+k^2_y)$. On the other hand for the
conduction band we found the trivial result: $\hat{H}(\mathcal{K})=\frac{k^2_x+k^2_y}{2m_e}$. 

Moreover, from the calculation of $\alpha$ ($\alpha=-1.31/m_0$), $\beta$
($\beta=-0.49/m_0$) and $m_e$ ($m_e=0.83m_0$) by interpolating the ab-initio
band structure (Fig.1) we found the mean value of the effective mass of the two nearly
degenerate excitons close to $\mu_{ex}=0.29$ $m_0$ ($m_0$ being the electron
mass).

\end{document}